# A new technique for the determination of the critical current density in superconducting films and flat samples

Conor McLoughlin[1,2], Pierre Bernstein[1], Yohann Thimont[1], J.Siejka[3,4]

[1] *CRISMAT-ENSICAEN (UMR-CNRS 6508), F14050 Caen, France*

[2] *National Centre for Plasma Science and Technology, School of Physical Sciences, Dublin City University, Glasnevin, Dublin 9, Ireland*

[3] *INSP Université Pierre et Marie Curie (UMR-CNRS 7588), F75251 Paris, France*

[4] *JAS Conseils, F75016 Paris, France*

**Abstract :** The determination of their critical current density in the whole range of the temperature below $T_c$ is of first importance to understand the physical processes occurring in superconducting films. We describe here a technique suitable for square films based on the measurement of the magnetic moment due to the currents persisting in the superconductor after the application of a perpendicular high magnetic field. Typically, with a SQUID magnetometer, the measurement time in the whole range of temperature with a 1K interval is of 2 hours only by this technique. An intriguing aspect of the obtained results is that they are much more accurate if the current lines are supposed to be circular than if we suppose, as suggested by theoretical considerations and magneto-optical observations that they have the sample symmetry.

**1. Introduction**

The classical techniques employed for determining the critical current density of thin films superconductors are the Bean method [1], that requires to record magnetic hysteresis loops and current-voltage measurements. Both techniques are time consuming. As a result, the measurements are carried out either in a restricted range of temperatures or at large intervals. This is especially detrimental in the case of the coated conductors, whose characterization is seldom carried out far below 77K, while their very large critical current density at low temperature is interesting for many applications. We have proposed for some time a technique suitable for square films [2,3]. Typically, with a SQUID magnetometer, the measurement time in the whole range of temperature with a 1K interval is of 2 hours only by this technique. In this contribution, we'll describe the measurement technique and we'll compare the results obtained with those gained from the use of other techniques. We'll discuss an intriguing aspect of the results that is that they are much more accurate if the current lines are supposed to be circular than if we suppose, as suggested by theoretical considerations and magneto-optical observations, that they have the sample symmetry.

Section 2 is devoted to the description of the measurements technique. We compare the results obtained to other measurements in section 3. In Section 4, the possible reasons that the circular symmetry and not that of the sample is relevant for the current lines in this type of measurements are discussed.

**2. Measurement technique**

The technique presented here is based on the measurement in self field, as a function of the temperature, of the magnetic moment **m** due to the currents persisting in a superconducting thin film after the application of a strong perpendicular magnetic field. In this section, we describe the measurement procedure and we give the relations permitting one to calculate the sheet critical current density of the sample, $J_{cr}^S$, i) if we suppose that the symmetry of the current lines is that of the sample and ii) if we suppose that they are circular.

The samples with a square shape are zero field cooled to the lowest measurements temperature. A high magnetic field, $\vec{B}_a$, is applied perpendicular to the sample plane and subsequently switched off. The same process is carried out with a reverse field. This procedure aims at establishing the same magnetic state in all the investigated samples and to generate screening currents. Then, the magnetic moment resulting from the persisting currents, is measured via a SQUID magnetometer or another technique, while increasing the temperature up to $T_c$. From classical electro-magnetism, the magnetic moment due to the currents circulating in a film with surface S takes the form :

$$1. \quad \vec{m} = \frac{1}{2} \iint_S \vec{r} \times \vec{J}^S dS$$

where $J^S$ is the sheet current density. For a an applied field larger than the field of complete penetration, $B_T$, the sheet density of the screening current can be regarded as equal to the critical value everywhere in the film and we can write [4]:

$$2. \quad J^S = J_{cr}^S$$

in Eq.(1). In the case of a 90 nm thick $YBa_2Cu_3O_{7-\delta}$ film cooled at 20K, Fig.1 shows that this condition is fulfilled if $B_a \geq 0.05T$, because **m**(T) does not depend on the amplitude of the applied field if $B_a$ is larger than this value. To ensure that it is always the case we routinely apply a 5T field. Then, considering that the current lines keep the sample symmetry, as suggested by theory [5] and magneto-optical observations [6], Eq.(1) results in :

$$3. \quad m = -J_{cr}^{S} \frac{w^3}{12}$$

and $J_{cr}^{S}(T)$ can be determined from **m**(T) and w. We have also investigated the possibility that the current lines have the circular shape shown in Fig.2. Eq.(1) takes then the form :

$$4. \quad J_{cr}^{S} = -\frac{24m}{\pi w^3 \left[1 + \frac{(\sqrt{2}-1)^3}{2}\right]}$$

In section III, we compare the $J_{cr}^{S}(T)$ obtained from measurements of **m**(T) with Eq.(3) and Eq.(4) to those obtained with other techniques.

## 3. Results

Th.Lécrevisse *et al.* [7] have carried out current-voltage measurements on a Superpower SCS4050 coated conductor at various temperatures. Fig.3 shows the sheet critical current density corresponding to these measurements as well as the $J_{cr}^{S}(T)$ obtained from **m**(T) measurements on the same sample using Eq.(3) and Eq.(4). Clearly, Eq.(4) gives results in better agreement with the transport measurements than Eq.(3). Table I compares the critical current density calculated for T=77K from magnetic moment measurements performed on various THEVA films with Eq.(3) and Eq.(4) to data provided by the manufacturer. For some films, an additional step consisting in etching the edges of the films was added to the fabrication process. Table I shows that for all the samples the critical current density calculated with Eq.(4) is a better approximation to THEVA's data than the $j_{cr}$ calculated with Eq.(3). We stress that the agreement is very good for all the etched films and fair for the non-etched samples. The reason is probably that the defects along the edges of the non etched films have some effect on the shape of the persisting current lines.

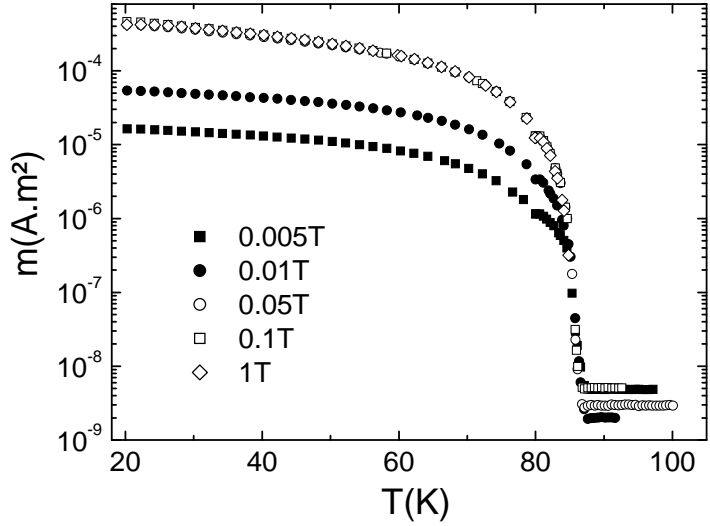

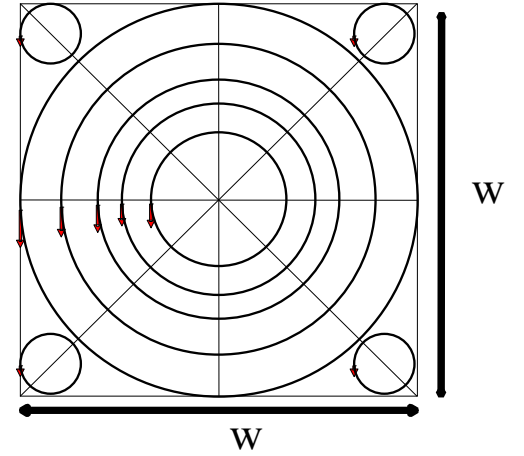

Fig.2: the circular current lines considered for deriving Eq.(4).

Fig.1 : magnetic moment of a 90nm thick YBCO film deposited on a SrTiO$_3$ substrate measured as a function of the temperature for various values of $B_a$

As a conclusion for this section, the comparison between our measurements and experimental results obtained by other techniques indicate that we can reliably calculate the sheet critical density of square thin films, as a function of the temperature, from the measurement of their magnetic moment in zero field, supposing that the persisting current lines are circular.

**Table I :** Comparison of the critical current density of THEVA films calculated with Eq.(3) and Eq.(4) to the values provided by the manufacturer.

| SAMPLE | Theva M 700 nm [140708] | Theva E 200 nm [280305 B] | Theva M 700 nm [071010 A] | Theva E 700 nm [161110 A] | Theva E 200 nm [151110 B] | Theva S 200 nm [081010 A] |
|---|---|---|---|---|---|---|
| $j_{cr}$ (A/m$^2$) square symmetry | 3.59 10$^{10}$ | 3.38 10$^{10}$ | 5.49 10$^{10}$ | 2.35 10$^{10}$ | 2.26 10$^{10}$ | 4.60 10$^{10}$ |
| $j_{cr}$ (A/m$^2$) circular symmetry | 2.2 10$^{10}$ | 2.1 10$^{10}$ | 3.4 10$^{10}$ | 1.5 10$^{10}$ | 1.4 10$^{10}$ | 2.85 10$^{10}$ |
| $j_{cr}$(A/m$^2$) THEVA | 2.5 10$^{10}$ | 2.8 10$^{10}$ | 3.3 10$^{10}$ | 1.6 10$^{10}$ | 1.5 10$^{10}$ | 2.6 10$^{10}$ |
| etched edges | no | no | yes | yes | yes | yes |

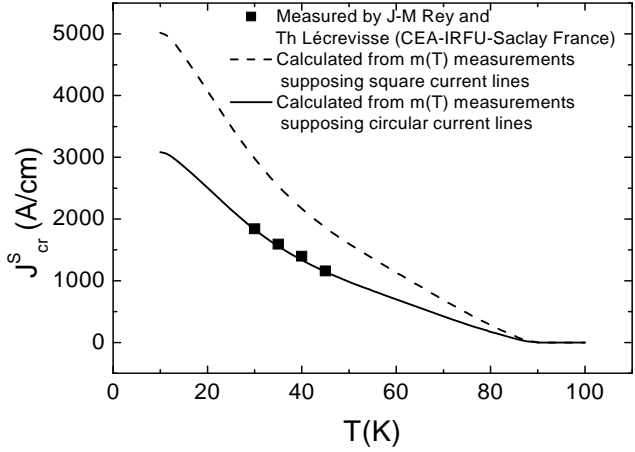 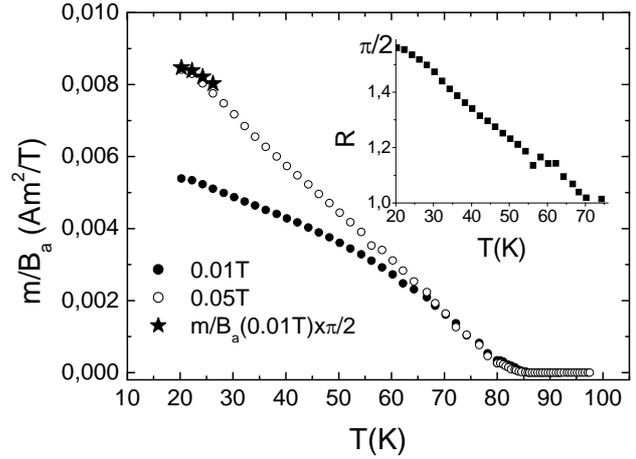

Fig.3 : Comparison between the sheet critical current density obtained from current-voltage measurements carried out on a SuperPower SCS4050 coated conductor and the values obtained from **m**(T) measurements using Eq.(3) and Eq.(4).

Fig.4 : $m/B_a$ (T) curves calculated for the sample in Fig.1 i) for $B_a$= 0.01T and ii) for $B_a$=0.05T. The inset shows the ratio R of $m/B_a$ at 0.05T to $m/B_a$ at $B_a$= 0.01T . The stars are the low temperatures $m/B_a$ of the $B_a$=0.01T curve multiplied by $\pi/2$ .

## 4. Discussion and conclusion

In this section, we discuss the reasons why $J_{cr}^S(T)$ can be calculated with accuracy if we suppose that the persisting current lines in square films are circular, while theory and magneto-optical observations suggest that they have the sample symmetry.

The model we have used is very naïve, since we have supposed that, for a large enough applied field, the sheet current density of the persisting current lines was equal to $J_{cr}^S$ everywhere in the superconducting film in the remnant state. As the applied field decreases from its maximal amplitude to zero, the flux distribution changes in the film, generating current lines that can have a direction opposite to that of the currents resulting from the previously applied field [8]. Due to this complicated current distribution, the calculation of $J_{cr}^S$ with circular current lines could result accidentally in a good approximation of the current density in the film, while the actual current lines have the square symmetry. However, Eqs. (3 and 4) show that for the same $J_{cr}^S$, circular current lines generate a larger **m** than square ones, while one expects that the existence of currents loops with opposite directions results in a lower magnetic moment. Another possibility arises from the comparison between measurements carried out with $B_a>B_T$ and $B_a<B_T$. Fig.4 shows the quantity $m/B_a$ calculated with the measurements reported in Fig.1, for $B_a= 0.01T$ and $B_a= 0.05T$. Above 67K, $m/B_a$ takes identical values for both curves. It has been established that in the parts of a film penetrated by the flux and the screening currents Eq.(2) is satisfied, although penetration is not complete [4], as it is the case when applying $B_a=0.01T$. Assuming that $B_T \approx 0.05T$, the similarity of the curves above 67K in Fig.4 suggests that i) the penetrated area does not change with the temperature and ii) that the modulation of $m/B_a$ is independent of the surface of this area and is due to that of $J_{cr}^S$ only. We stress that, of course, this is valid only if $B_a \leq B_T$. Below 67K, the $m/B_a$ curves have different shapes. If we suppose that the symmetry of the current lines is circular for $B_a= 0.05T$ in the whole temperature range, this suggests that it is the case above 67K only if the applied field is equal to 0.01T. For a completely penetrated square sample, from Eqs.(3 and 4), we have :

$$5. \quad \frac{m_c}{m_r} \approx \frac{\pi}{2}$$

where $m_c$ and $m_r$ are the magnetic moments resulting, for the same $J_{cr}^S$, from circular and square current lines, respectively. The same relation should hold for $m/B_a$, if this quantity depends on $J_{cr}^S$ only. The inset in Fig.4 shows the ratio of $m/B_a$ for $B_a= 0.05T$ to $m/B_a$ for $B_a=0.01T$. This ratio increases from 1 above 67K to $\pi/2$ in the 20K-25K range. This result suggests strongly that, while the current lines are rectangular at low temperature when the applied field is equal to 0.01T, a transition towards circular current lines occurs in the 25K-67K domain. More generally one can suspect that the shape of the current lines is rectangular for $B_a<B_T$, at least at low temperature, while it is circular in the whole temperature range for $B_a>B_T$. The probable reason for these changes in the current lines symmetry is that circular current lines yields a reduction in the vortex density and, as a result, in the inter-vortex repulsive energy since they generate a larger **m**. Work is in progress to make these aspects clear.

## 5. Acknowledgments:

C. McLoughlin acknowledges support from the Irish Research Council for Science, Engineering, and Technology through the INSPIRE Marie Curie cofund fellowship scheme.

J.Siejka is grateful to doctor R. Semerad from Theva GmbH for samples preparation.